\documentclass[%
 reprint,
superscriptaddress,
nofootinbib,
 amsmath,amssymb,
 aps,
]{revtex4-2}

\usepackage{graphicx}
\usepackage{dcolumn}
\usepackage{bm}
\usepackage{amsmath}
\usepackage{ amssymb }
\usepackage{xcolor}
\usepackage[version=4]{mhchem}
\usepackage{listings}
\usepackage[utf8]{inputenc}

\usepackage{float}
\usepackage{placeins} 

\usepackage{hyperref}

\newcommand\norm[1]{\left\lVert#1\right\rVert}
\newcommand{\cycle}{t}

\newcommand{\R}{\mathbf{r}}
\newcommand{\Z}{Z}
\newcommand{\e}{E}

\newcommand{\natoms}{N}

\newcommand{\features}{\mathbf{x}_{\theta}}
\newcommand{\featuresi}{\mathbf{x}_{\theta, i}}

\newcommand{\epred}{E_{\theta}}
\newcommand{\epredi}{E_{\theta, i}}
\newcommand{\fpred}{\mathbf{f}_{\theta}}
\newcommand{\spred}{\mathbf{s}_{\theta}}
\newcommand{\edft}{\e_{\text{DFT}}}
\newcommand{\fdft}{\mathbf{f}_{\text{DFT}}}
\newcommand{\sdft}{\mathbf{s}_{\text{DFT}}}
\newcommand{\emean}{\overline{E}_{\theta}}
\newcommand{\fmean}{\overline{\mathbf{f}}_{\theta}}
\newcommand{\smean}{\overline{\mathbf{s}}_{\theta}}

\newcommand{\euncert}{\sigma_{E, \theta}}
\newcommand{\funcert}{\boldsymbol{\sigma}_{f, \theta}}
\newcommand{\suncert}{\boldsymbol{\sigma}_{s, \theta}}
\newcommand{\uncert}{U_{\theta}}
\newcommand{\pes}{E(\Z_{1}, ..., \Z_{\natoms}, \R_{1}, ..., \R_{\natoms})}

\newcommand{\score}{L_{\theta, j}}
\newcommand{\noise}{\boldsymbol{\epsilon}}
\newcommand{\noiset}{\noise_{t}}
\newcommand{\noisedamping}{\omega_{\epsilon}}

\begin{document}

\preprint{APS/123-QED}

\title{Accelerating crystal structure search through active learning with neural networks for rapid relaxations}

\author{Stefaan S. P. Hessmann}
\email{stefaan.hessmann@tu-berlin.de}
\affiliation{Machine Learning Group, Technische Universit\"at Berlin, 10587 Berlin, Germany}
\affiliation{Berlin Institute for the Foundations of Learning and Data, 10587 Berlin, Germany}

\author{Kristof T. Schütt}
\affiliation{Machine Learning Group, Technische Universit\"at Berlin, 10587 Berlin, Germany}
\affiliation{Berlin Institute for the Foundations of Learning and Data, 10587 Berlin, Germany}

\author{Niklas W. A. Gebauer}
\affiliation{Machine Learning Group, Technische Universit\"at Berlin, 10587 Berlin, Germany}
\affiliation{Berlin Institute for the Foundations of Learning and Data, 10587 Berlin, Germany}
\affiliation{BASLEARN – TU Berlin/BASF Joint Lab for Machine Learning, Technische Universit\"at Berlin, 10587 Berlin, Germany}

\author{Michael Gastegger}
\affiliation{Machine Learning Group, Technische Universit\"at Berlin, 10587 Berlin, Germany}
\affiliation{BASLEARN – TU Berlin/BASF Joint Lab for Machine Learning, Technische Universit\"at Berlin, 10587 Berlin, Germany}

\author{Tamio Oguchi}
\affiliation{Center for Spintronics Research Network, Osaka University, 1-3 Machikaneyama, Toyonaka, Osaka 560-8531, Japan}

\author{Tomoki Yamashita}
\affiliation{Department of Electrical, Electronics and Information Engineering, Nagaoka University of Technology, 1603-1 Kamitomioka-machi, Nagaoka, Niigata, 940-2188, Japan}

\date{\today}

\begin{abstract}
Global optimization of crystal compositions is a significant yet computationally intensive method to identify stable structures within chemical space.
The specific physical properties linked to a three-dimensional atomic arrangement make this an essential task in the development of new materials.
We present a method that efficiently uses active learning of neural network force fields for structure relaxation, minimizing the required number of steps in the process.
This is achieved by neural network force fields equipped with uncertainty estimation, which iteratively guide a pool of randomly generated candidates towards their respective local minima.
Using this approach, we are able to effectively identify the most promising candidates for further evaluation using density functional theory (DFT).
Our method not only reliably reduces computational costs by up to two orders of magnitude across the benchmark systems \ce{Si16}, \ce{Na8Cl8}, \ce{Ga8As8} and \ce{Al4O6}, but also excels in finding the most stable minimum for the unseen, more complex systems \ce{Si46} and \ce{Al16O24}.
Moreover, we demonstrate at the example of \ce{Si16} that our method can find multiple relevant local minima while only adding minor computational effort.

\end{abstract}

\maketitle


\section{\label{sec:introduction}Introduction}
Novel crystal structures have allowed significant improvements across various research fields, for example, in the discovery of solar cells \cite{green2014emergence, correa2017rapid}, catalysts \cite{norskov2009towards, greeley2006computational}, superconductors \cite{bednorz1986possible}, hardware components \cite{liu2020two, de2021materials} and batteries \cite{mizushima1980lixcoo2, ceder1998identification}.
For a single crystal composition, a vast number of stable structures can be found with each having a unique set of physical properties, such as electrical conductivity, thermal conductivity, magnetism, and optical behavior.
Stable crystal structures are local minima on the potential energy surface (PES) of the respective composition, and the number of possible stable structures increases exponentially with the number of atoms per cell \cite{stillinger1999exponential}.
As a result of the wide variety of possible stable structures and their physical properties, crystal structure search is an important challenge, with global energy optimization being the most fundamental task. 

Many computational methods have been developed for crystal structure search and, in particular, global optimization.
These approaches usually start from a pool of general candidate structures that are relaxed to local minima in the PES.
During structure relaxation, atom positions and lattice vectors are adjusted such that a structure is moved towards its respective stable structure, resulting in a trajectory of intermediate structures for which at every step the energy, forces, and stress need to be evaluated.
To reliably compute physical properties and identify all possible local minima in the PES of a crystal structure composition, accurate electronic structure computations such as density functional theory (DFT) are inevitable.
The simplest crystal structure search method is ab initio random structure search (AIRSS) \cite{pickard2006high, pickard2007structure, pickard2011ab}, where the structures of the candidate pool are randomly selected for evaluation with structure relaxation.
However, the calculation of physical properties for all intermediate structures is computationally expensive, making the high-throughput virtual screening (HTVS) of many candidate structures infeasible for large systems with many local minima.
To reduce the need for DFT calculations required for structure relaxation, the number of trajectories can be reduced with iterative machine learning methods.
Only the most promising candidates will be selected for relaxation, for example, using Bayesian optimization \cite{yamashita2018crystal, sato2020adjusting}.
Alternatively, the computational cost can be reduced by transforming the structures in the candidate pool such that the probability for finding low-energy minima is increased.
These approaches include evolutionary algorithms \cite{oganov2006crystal, oganov2011evolutionary, lyakhov2013new}, basin hopping \cite{wales1997global}, minima hopping \cite{goedecker2004minima, amsler2010crystal}, particle swarm optimization \cite{wang2010crystal, zhang2017computer}, and simulated annealing \cite{kirkpatrick1983optimization, pannetier1990prediction}.
In addition, a combination of candidate selection and transformation of the candidate pool has been used for large scale discovery of materials \cite{merchant2023scaling}.
However, all these methods rely on computationally expensive full relaxation trajectories based on ab initio methods such as DFT, in order to find stable structures starting from general, usually non-stable candidates. 
As long as the full structure relaxation trajectories need to be computed, the potential for reducing computationally expensive DFT steps remains limited.
To address this problem, a small number of promising methods have been developed that stop the relaxation of the structure in intermediate stages, using the energies, forces, and stress of those intermediate structures, combined with a scoring function \cite{terayama2018fine, yamashita2022improvement}.
On the other hand, machine learning force fields (MLFFs) \cite{unke2021machine, keith2021combining, unke2024biomolecular, chmiela2023accurate} such as neural network models \cite{schnet, schutt2018schnet, painn, dtnn, smith2017ani1_potential, dimenet, nequip, satorras2021n, so3krates, batatia2022mace, allegro, tensorfieldnet} or kernel methods \cite{chmiela2017machine, sgdml}, have shown to accurately learn the PES of molecules and materials based on a set of structures and labels for physical properties, while reducing computational cost by multiple orders of magnitude.
When trained on a sufficiently large dataset, machine learning models have proven to be useful tools for structure relaxation \cite{physnet, spookynet, kahouli2024morered}. 
However, in many applications, including crystal structure search, suitable datasets for training the MLFFs are often not available and need to be created.
To assemble such data sets efficiently while avoiding faulty MLFF predictions for structures that are very different from the training data, active learning has been successfully applied to MLFF training \cite{vandermause2020fly, yang2021machine, gastegger2017machine}. This includes training moment tensor potentials \cite{shapeev2016moment} for HTVS experiments \cite{gubaev2019accelerating} and in evolutionary algorithms \cite{podryabinkin2019accelerating}.
Moreover, neural network MLFFs have been trained with active learning for the accelerated discovery of transition metals \cite{nandy2021computational} and organic molecular crystals \cite{butler2024machine}.
There also exist some generative approaches that do not rely on the evaluation of candidate structures but instead use diffusion models \cite{mattergen} that are trained on large databases of stable materials such as MaterialsProject \cite{Jain2013}, the open quantum materials database (OQMD)\cite{saal2013materials}, AFLOWLIB.ORG \cite{curtarolo2012aflowlib}, and NOMAD \cite{scheidgen2023nomad}.

In this work, we propose an iterative approach to global optimization in crystal structure search based on active learning and structure relaxations of large candidate pools with neural network MLFF ensembles.
The key component of our method is a neural network ensemble that accelerates structure relaxations, selects new training data points towards the region of interest, finds low-energy clusters in the candidate pool, and finally provides a stopping criterion to measure convergence.
Using ensembles of MLFFs to measure the uncertainty of predictions, which is required for active learning, makes our method agnostic to the type of MLFF.
The only requirement is that the models allow to derive a feature representation, as is the case for all commonly used state-of-the-art models.
Especially with the increasing availability of pre-trained models and datasets \cite{matproj, merchant2023scaling}, a flexible choice of model architectures allows one to start from trained models and, for this reason, possibly further reduce the need for expensive DFT calculations during active learning.
Furthermore, our method enables straight-forward parallelization of computationally demanding tasks, making it efficient for use on high-performance computing (HPCs).

We evaluate our method for global optimization of \ce{Si16}, \ce{Na8Cl8}, \ce{Ga8As8}, and \ce{Al4O6}.
Here, we reduce the computational effort by up to two orders of magnitude compared to approaches without active learning.
We show that the method is capable of finding multiple relevant local minima in the low-energy region of \ce{Si16} without a large increase in computational effort, compared to only finding the global energy minimum.
Finally, we demonstrate the transferability of the neural network ensembles to relatively larger systems by training the models only on the smaller structures of \ce{Si16} and \ce{Al4O6} and using them for structure search of \ce{Si46} and \ce{Al16O24}, respectively.
This additionally diminishes the computational expense of DFT calculations throughout the active learning phase.

\begin{figure}
    \centering
    \includegraphics[width=0.46\textwidth]{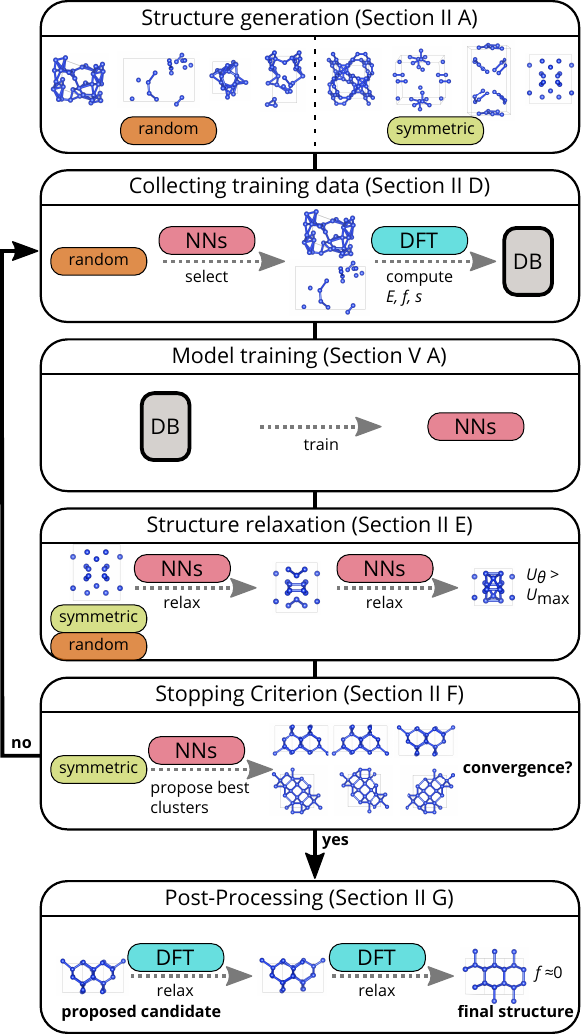}
    \caption{Schematic overview of the method. Initially, a pool of candidate structures is generated based on random or symmetric generation. Next, the active learning cycle is started by the selection of promising datapoints and by training the neural network ensembles. These are then used to accelerate structure relaxations for all structures in the candidate pool. At the end of the active learning cycle, the neural network models propose the most promising clusters of low-energy structures and a stopping criterion is applied. In the case of convergence, the most promising structures are further validated through structure relaxation with DFT.}
    \label{fig:overview}
\end{figure}

\section{Results}
Our approach is initialized with the generation of a pool of candidates, which are typically still far from their respective local minimum in the PES, and subsequently applies multiple rounds of active learning, which we will refer to as cycles, to find the global energy minimum.
A schematic overview of the different stages of our method is shown in Figure \ref{fig:overview}.
Each cycle begins by sampling new training data based on a scoring function to specifically target the low-energy region of the PES.
In the next step, we perform DFT computations to obtain energies, forces and stress for the sampled structures and update the training database accordingly.
We then train an ensemble of neural network MLFFs on the updated training data and use the trained models for structure relaxation of all structures in the candidate pool until an uncertainty criterion is triggered for every structure, respectively.
Finally, we define a convergence criterion based on low-energy clusters in the optimized candidate pool that are obtained by using the MLFFs. 
The cycle is repeated until the convergence criterion is fulfilled, and the low-energy structures are selected for further post-processing steps and validation with DFT.
The following sections will provide a detailed description of the different stages that are used in our method.

\subsection{\label{subsec:initial_structures}Initial structure generation}

\begin{figure}[!t]
    \centering
    \includegraphics[width=0.48\textwidth]{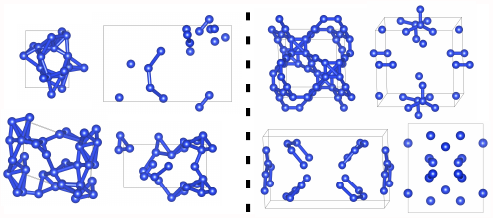}
    \caption{Renders of \ce{Si16} candidate structures. Structures on the left are generated with random generation and structures on the right are generated with symmetric generation.}
    \label{fig:initial_structures}
\end{figure}

Structure generation can either be done naively by generating an empty cell and filling it with atoms at random positions, which we call \emph{random generation}, or by sampling a space group and generating a structure that satisfies the symmetry sampled with the use of PyXtal \cite{pyxtal}, which we call \emph{symmetric generation}.
For both strategies, we use the structure generator as implemented in CrySPY \cite{yamashita2021cryspy}, and we show example structures for each method that have been rendered with VESTA \cite{momma2008vesta} in Figure \ref{fig:initial_structures}.
To keep the initial candidates within a reasonable range, minimum interatomic distances and ranges for cell volumes are applied (see Section \ref{subsec:distance_constraints}).

We observe that candidates obtained by symmetric generation generally require fewer steps during structure optimization and are therefore well suited for classical crystal structure search algorithms.
However, we also find that highly symmetric crystal structures provide challenging data points for training MLFFs, which will often lead to inaccurate predictions compared to models trained on structures without symmetries (see Supplementary Materials \ref{app:mlff_symmetry}).
In contrast, training MLFFs on structures that are created by random generation or by adding perturbations to symmetric structures \cite{cheon2020dataset} significantly improves accuracy.
This may be due to the fact that structural symmetry often leads to repeated local geometries within a single cell, causing some local atomic neighborhoods to be rotated copies of other local atomic neighborhoods, thus reducing variance in the training data and over-representing some local environments compared to others.
The longer optimization trajectories of randomly generated structures make them poor candidates for classical crystal structure search algorithms. 
Furthermore, some symmetries might be hard to find based solely on fully random candidates.

As our method requires good candidates for discovering local minima through structure relaxation and for training neural network force fields, we generate candidates with both random and symmetric generation to bridge the gap.
We use a split pool with randomly generated structures for training the MLFFs and symmetric structures that provide promising candidates to find local minima in the PES.

\subsection{\label{subsec:neural_networks}Neural network models}
We use an SO(3)-equivariant neural network architecture as implemented in SchNetPack \cite{schutt2023schnetpack} to learn the PES $\pes$ based on nuclear charges $\Z_{i} \in \mathbb{N}$ and positions $\R_{i} \in \mathbb{Z}^{3}$ with the atom index $i$.
The neural network can be divided into two main components.
First, for every atom $i$ in the cell of $N$ atoms, a message passing architecture computes an $f$-dimensional, roto-translational invariant feature representation $\featuresi \in \mathbb{R}^{f}$, based on its local atomic environment.
Subsequently, a multilayer perceptron reduces the dimensionality of the feature vectors $\featuresi$ to a scalar $\epredi$, that is, the so-called local energy contribution, and aggregates these energy contributions of the local environments to obtain the total energy of the cell $\epred = \sum_{i=1}^{N} \epredi$.
Response properties such as forces $\fpred$ and stress $\spred$ are calculated as derivatives of the energy prediction with respect to atomic positions and cell parameters, respectively.

For active learning, it is crucial to find promising new datapoints for labeling and to filter out predictions with low accuracy, which is usually done based on uncertainty estimates of model predictions.
Because uncertainty estimates cannot be easily derived from neural networks, we train an ensemble of independent models on the same data set.
The prediction of physical properties is easily obtained from the ensemble by using the mean predictions $\emean$, $\fmean$ and $\smean$ and an uncertainty estimate can be defined based on the standard deviations $\euncert$, $\funcert$ and $\suncert$.
Due to the fact that ensembles are a general concept, they are invariant against the model architecture, allowing every MLFF to be used within this active learning setup.

A detailed description of the model architecture and the training settings can be found in Section~\ref{subsec:training}.

\subsection{\label{subsec:clustering}Clustering}
In a large pool of candidates, many structures will, after a sufficiently large number of structure relaxation steps, converge to the same local minimum of the PES.
To avoid costly redundant reference calculations and to identify local energy minima, clustering methods can find similar structures in the pool of candidates.

However, comparing the similarity between structures with periodic boundary condition (PBC) is not straightforward.
For example, duplicating the cell into any direction does not change any physical properties of the system, and small positional changes of atoms near the edges of the cell can result in displacement of the atom towards the opposite side of the cell.
To address this, we map a cell with an arbitrary number of atom types $\Z_{i}$ and positions $\R_{i}$ to a descriptor in a feature space $\features \in \mathbb{R}^{f}$ with a fixed dimension.
We constrain our descriptor to invariances against translation, rotation, and permutation of atoms, as well as to symmetry constraints due to the periodicity of the system.
All these constraints, except for invariance to permutation of atoms, are fulfilled by the  roto-translational invariant feature representations $\featuresi$ of the trained neural network MLFFs.
Invariance to permutation of atoms is achieved by applying a mean pooling over the atomic dimension $i$, which results in a global descriptor $\features$:

\begin{equation}
    \features = \frac{1}{N} \sum_{i=1}^{N} \featuresi
    \label{eq:representation}.
\end{equation}

After computing the global descriptors for all candidates in the pool, the descriptors are standardized to have the same mean and standard deviation in every dimension.
In principle, the standardized global descriptors could be used with any clustering algorithm.
In our problem setting, the number of local minima is unknown and many structures may be too far from their stable structure to be clustered.
Therefore, we use the density-based clustering algorithm DBSCAN \cite{ester1996density} as implemented in scikit-learn \cite{scikit-learn}, which handles outliers well and finds the required number of clusters automatically.

\subsection{\label{subsec:single_point}Collecting training data}
For the first cycle, no knowledge about the candidate structures is available, except for cell shape, atom types and positions.
For this reason, the first batch of candidate structures is selected at random.
For all subsequent cycles, the property predictions $\emean$, $\fmean$ and $\smean$ of the trained ensembles, as well as their uncertainty estimates $\euncert$, $\funcert$ and $\suncert$ can be used to improve the selection of promising candidates.
In standard active learning algorithms, the database of unlabeled data remains fixed and the most promising datapoints for labeling are selected purely on the uncertainty estimates of the most recent, trained model.
However, in our approach, the unlabeled data, i.e. the pool of candidates, are updated during every active learning cycle by using the trained MLFFs for structure relaxation.
Because the structure relaxation of a single input structure is usually performed until the uncertainty criterion is violated, the unlabeled structures in the pool of candidates generally are at the uncertainty threshold, making them good candidates for labeling.
To efficiently guide our algorithm towards finding low-energy minima, we select structures with a high potential for low relaxation energies.
Although an abundance of selection strategies could be used to select the next batch of candidates, we use a fairly simple criterion based on the predicted energies $\emean$ and forces $\fmean$.
Our criterion is inspired by the LAQA \cite{terayama2018fine} selection method and assigns high scores $\score$ to structures with low-energies $\emean$ and high forces $\fmean$ according to equation \eqref{eq:selection_criterion}.

\begin{equation}
    \score = -\overline{E}_{\theta, j} + \omega_{f} \cdot \max(||\overline{\mathbf{f}}_{\theta, j}||_2)
    \label{eq:selection_criterion}
\end{equation}
The weighting parameter $\omega_{f}$ can be used to find a trade-off between the exploitation of low-energy structures and the exploration of structures with a high potential for further optimization due to their large forces.
After computing the selection scores for all structures in the candidate pool, we rank all structures accordingly.
Because similar structures are assigned to similar selection scores, we detect clusters of similar structures with clustering, as described in Section \ref{subsec:clustering} and only allow the structure with the highest score $\score$ per cluster for selection.
Now, the batch of candidates with the highest scores is selected for evaluation with single point DFT-calculations.

As already mentioned in Section \ref{subsec:initial_structures}, highly symmetric structures lead to a possible reduction in the prediction accuracy of the trained MLFFs.
Over the course of multiple cycles, structure relaxation will also increase the symmetry in the training structures that have been created through random generation and therefore make them less valuable training data points. 
To prevent this, we break the symmetry of the selected structures by adding Gaussian noise $\noise$ to the positions $\R_{i}$:

\begin{equation}
    \R_{i, \epsilon} = \R_{i} + \mathbf{N}(0, \noise)
    \label{eq:noisy_positions}.
\end{equation}
But because large perturbations disturb the stability of selected candidates, small values of inter-atomic forces $\fdft$ are less likely, depending on the noise level.
To allow for smaller forces in the data set during subsequent cycles $\cycle$ of the method, the noise level is scheduled by a damping factor $\noisedamping$ according to equation \eqref{eq:scheduled_noise}.

\begin{equation}
    \noiset = \noise \cdot \noisedamping^{\cycle}
    \label{eq:scheduled_noise}
\end{equation}

Finally, the set of perturbed candidates is evaluated with DFT to compute the energies $\edft$, the forces $\fdft$, and the stress $\sdft$.
Details on the reference computations can be found in Section~\ref{subsec:dft_settings}.

\subsection{\label{subsec:pool_relaxation}Structure relaxation}
After retraining the MLFFs on the updated dataset, they are used for structure relaxation of the candidate pool.
To reduce computational cost, we replace expensive DFT computations with the fast MLFF predictions for energy, forces, and stress.
We avoid relaxation steps in regions of the PES with large prediction errors by using an uncertainty criterion $U_{\theta}$ based on the mean force predictions $\fmean$ and the standard deviation of the prediction $\funcert$:

\begin{equation}
    \uncert = \frac{\max(||\funcert||_2)}{\max(||\fmean||_2)}
    \label{eq:uncertainty_crit}.
\end{equation}
During structure relaxation, initial structures are transformed according to the derivatives of the PES with respect to positions and cell parameters, thus, the forces and stress values decrease during the process.
Structures close to their local minima require very accurate predictions, whereas less accurate predictions are often sufficient for structures far from the local minima.
To account for this behavior, we use a relative uncertainty criterion that allows larger absolute errors for structures with large force predictions.
Although the stress computations $\smean$ also matter during structure relaxation with PBC, they are not considered in the uncertainty criterion.
This is because the stress computations are directly dependent on accurate force predictions.
Furthermore, we observe that stress predictions converge mainly during the early steps of structure relaxation, while forces are often optimized later in the process.
Therefore, a relative uncertainty criterion for stress predictions would often yield very large values, although the models still predict meaningful values for the forces.

We optimize every structure in the pool of candidates until either the optimization has converged, the maximum number of steps has been reached, or the uncertainty criterion exceeds the threshold $U_{\text{max}}$.
Afterwards, the original structure in the candidate pool is replaced with its optimized version.
This drives the pool of candidates closer to their local minima with limited computational effort, while large inaccuracies of the neural network predictions are prevented by the applied uncertainty criterion.

\subsection{\label{subsec:dft_relaxation}Stopping criterion}
After a number of cycles, structure relaxation with the MLFFs has decreased the energies of structures in the candidate pool to such an extent that they are close to their respective local minima in the PES.
In a sufficiently large candidate pool, many structures converge to the same local minima so that they can be detected through clustering as it is described in Section~\ref{subsec:clustering}.

To measure the convergence of our algorithm, we utilize this observation and compare low-energy clusters of the current candidate pool to the low-energy clusters that are found in candidate pools of previous cycles.
That is, we track the candidate pools of the previous $n_{\text{cycle}}$ cycles and use the clustering algorithm with the most recent MLFF to obtain the respective clusters.
We then define the cluster energies $\emean^{l, m}$, as the lowest energy that is present in one cluster as predicted by the MLFF and sort the clusters accordingly, with $l$ being the sorted cluster idx and $m$ the cycle index of the candidate pool.
Based on the cluster energies we apply the convergence criterion $\emean^{l, c}-\emean^{l, c-m} < \epsilon$ with $l = 1, ..., k$, $m = 1, ..., n_{\text{cycle}}$ at cycle $c$.
If no improvement greater than $\epsilon$ is obtained for the $k$ most stable clusters during the selected number of cycles, the method is terminated.
However, if the convergence criterion is not met, the next cycle is started by selecting new structures for single point computations.
The number of stable clusters $k$, the number of previous cycles to track $n_{cycle}$ and the maximum energy difference $\epsilon$ are hyperparameters of our stopping criterion.

\subsection{\label{subsec:postprocessing}Post-processing}
Once the stopping criterion has been triggered, we validate the top $k$ stable structures.
However, depending on the chemical composition, clustering can be inaccurate for some cases such that a local minimum can be split into sub-clusters, remains undetected, or two local minima could be merged into a single cluster resulting in one minimum not being selected.
For example in the compositions \ce{Si} or \ce{GaAs}, the global energy minimum and the second lowest local minimum are structurally very similar and only have a minor energy difference of about 0.01 eV/atom, and we observe that their clusters can falsely be merged into one cluster.
But because the number of selected low-energy clusters is small, these mistakes can easily be detected.
Merged clusters can be found, for example, by visually inspecting the clusters for dissimilar structures, using structure matching tools with a threshold value or by measuring energy differences within the cluster.
Furthermore, by checking if the structures with the lowest energy predictions in the candidate pool are part of the selected low-energy clusters, we can account for undetected clusters or single low-energy structures.

Finally, after selecting the low-energy structures for validation, we confirm the energies of the respective local minima by structure relaxation with DFT.
Structure relaxation with DFT involves multiple single point DFT calculations, introducing significantly larger computational cost than the calculation of physical properties, as it is done in Section \ref{subsec:single_point}.

\section{Simulations\label{sec:simulations}}
\subsection{\label{subsec:global_minimum}Global energy minimum}

To evaluate the performance of our approach, we select a number of benchmark systems with well-known low-energy minima.
We run simulations on \ce{Si16}, \ce{Na8Cl8}, \ce{Ga8As8} and \ce{Al4O6} and compare our findings with the baseline methods random search, Bayesian optimization, and LAQA as reported in \cite{terayama2018fine}.
The global energy minima that we target are the structures MP-149 for \ce{Si16}, MP-2534 for \ce{Ga8As8}, MP-22862 for \ce{Na8Cl8} and MP-1143 for \ce{Al4O6} that we obtain from the materials explorer of MaterialsProject \cite{materialsexplorer, Jain2013, matproj}.

For every system, we generate a candidate pool containing 30 starting structures for all possible space groups that are valid for the respective composition as well as 1000 randomly generated structures.
Only structures from the pool of initial candidates without any symmetry constraints are selected as data points for training.
A detailed list of parameters for the DFT computations is shown in Section~\ref{subsec:dft_settings} and the hyperparameters for the neural network ensembles are shown in Section~\ref{subsec:training}.
For the stopping criterion, we target the $k=3$ lowest energy clusters and track the $n_{\text{cycle}}=1$ last cycles.
All experiments are performed until the stopping criterion terminates the simulation.

\begin{figure}
    \centering
    \includegraphics[width=0.5\textwidth]{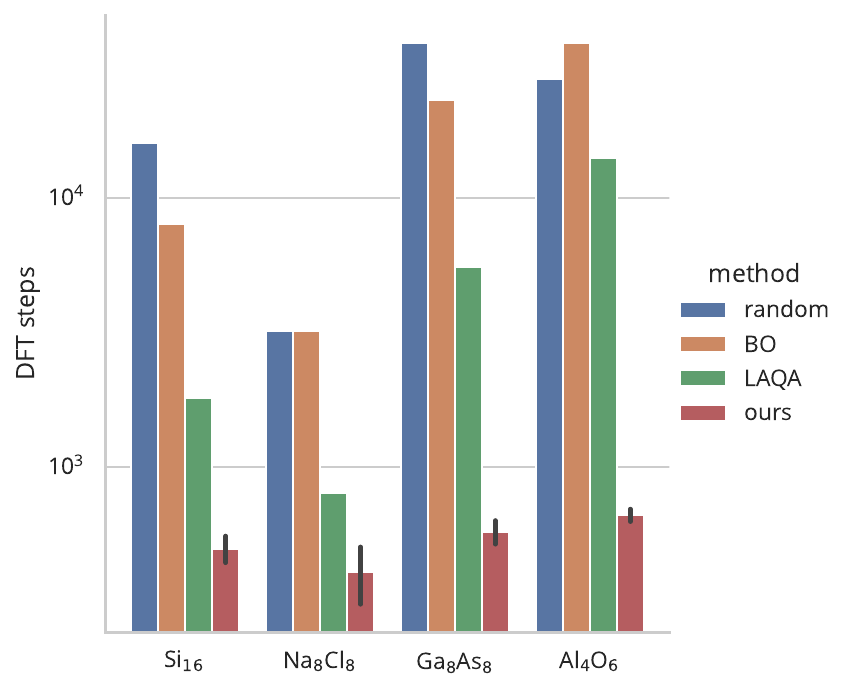} \quad
    \caption{Efficiency of various methods in identifying global energy minima. The bar plot shows a comparison of DFT single point evaluations required for finding the global energy minimum of different chemical compositions for random search, Bayesian optimization (BO), look ahead based on quadratic approximation (LAQA), and our method. Error bars show the standard deviation over a set of 3 simulations with equal settings.}
    \label{fig:results}
\end{figure}

To account for the stochasticity of our method, we perform three simulations for every composition and evaluate the total number of single-point DFT computations as well as the standard deviation which is shown in the error bars.
Our analysis includes the number of computations that are used for labeling of training data, as well as the number of steps for validating the predicted local minima.
Furthermore, it also includes unconverged DFT calculations that may occur due to exceeding the timeout limit or because the maximum number of convergence steps is reached.
The results of the experiments are shown in Figure \ref{fig:results}.

For all systems tested, our algorithm requires a similar number of 400-700 single-point DFT evaluations until convergence.
This stands in strong contrast to AIRSS, BO and LAQA, where the number of DFT computations highly depends on the composition of the selected material and, therefore, also on the number of possible local minima in the PES as well as the likelihood for the global energy minimum.
Please note that we run our method until the stopping criterion is triggered, while the other methods only run until the targeted global minimum structure is observed.
Since this cannot be done for compositions where the global minimum is unknown, the number of reference calculations required with the baseline methods should be interpreted as optimistic lower bounds.
It remains unclear how many additional steps would be required until their stopping criterion is reached.
In comparison to the baselines, we reduce the number of expensive DFT evaluations by up to two orders of magnitude for the most challenging systems \ce{Al4O6} and \ce{Ga8As8}.
We provide a detailed runtime analysis in the Supplementary Materials~\ref{app:runtime}, where we show that the additional computational cost of our approach due to training and utilizing the neural netwrok ensembles is negligible compared to the DFT calculations.
Except for one of the three simulations of \ce{Si16}, our method reliably finds the global minimum based on clustering the candidate pool and without further post-processing steps.
In this one case, where the clustering did not provide the global energy minimum, the two lowest energy clusters had been merged due to their similarity and could easily be separated by hand.

\begin{figure*}[]
  \centering
  \includegraphics[width=\textwidth]{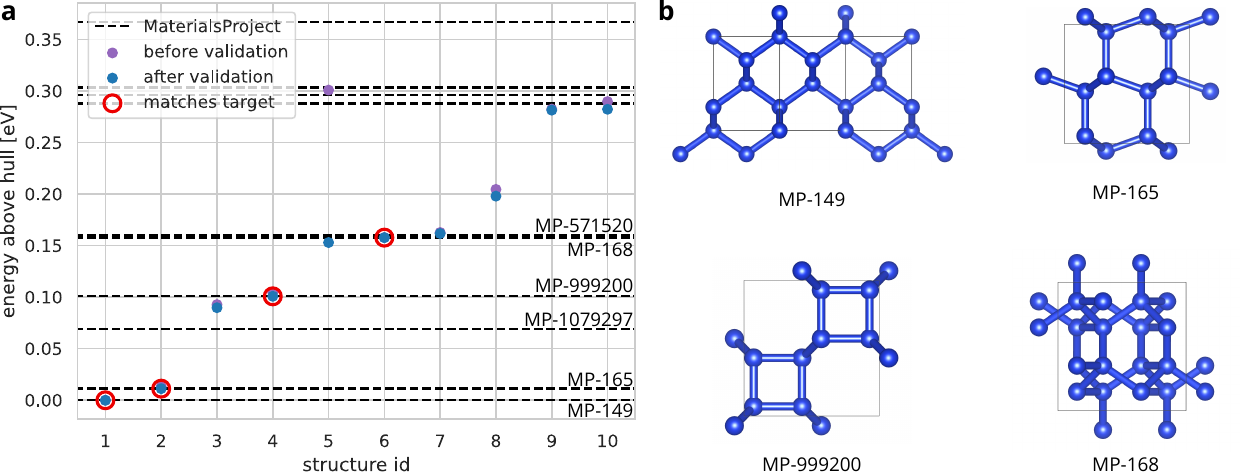}
  \caption{
  \textbf{a} DFT energies of the lowest 10 local minimum structures of \ce{Si16}.
    Purple dots represent the energies of structures as they are proposed by our method after convergence and the blue dots show the energies of those structures after validation through structure relaxation with DFT. 
    The dashed lines mark energy levels of known \ce{Si16} structures that have been reported in MaterialsProject and red circles are drawn for structures where our method matches one of the target structures.
    \textbf{b} Renders of \ce{Si16} local minima. Renders of structures that are found by our method and can be matched to structures in MaterialsProject (see red circles).
  }
  \label{fig:top10relaxations}
\end{figure*}
\FloatBarrier
\subsection{\label{subsec:multi_search}Multiple low-energy structures}

In many cases, not only the structure at the global energy minimum, but also other stable low-energy structures are of interest.
Because our method performs structure optimization on all structures in the candidate pool, other local minima can also be found without large additional computational efforts.
To investigate this, we continue with one simulation of \ce{Si16} from the previous section but change the parameter $k$ that determines the number of low-energy clusters for monitoring from previously 3 to 10.
Once the stopping criterion is triggered, we validate the 10 most promising low-energy clusters that are proposed by our algorithm through structure relaxation with DFT.
We find that for this particular simulation, the stricter convergence criterion does not have any effect, so that the simulation does not perform any additional active learning cycles.
This concludes a total number of 500 DFT calculations during the active learning cycles, 27 DFT steps to validate the lowest 3 energy clusters and an additional 121 DFT calculations are required to validate the low-energy minima 4-10.
Figure~\ref{fig:top10relaxations} compares the energies of structures proposed by our simulation before and after validation with the top 10 lowest energy structures for \ce{Si16} that are reported in MaterialsProject.

We observe that our simulation finds 10 local minima in the PES of \ce{Si16} with energies that are on average lower than the 10 lowest energy structures of MaterialsProject.
The structure with the highest energy is still below the 7th lowest energy minimum of the target structures, and we are able to find 4 out of the 5 lowest energy structures of MaterialsProject.
The other 6 structures found by our simulation are stable with regard to the DFT validation, but are not reported in MaterialsProject.
Figure \ref{fig:top10relaxations}b displays renders of our structures that correspond to those of MaterialsProject, and the full set of structures can be found in the Supplementary Materials~\ref{app:si_minima}.
Apart from one outlier, all structures proposed by our method are very close to the local minima based on validation through structure relaxation with DFT.
To validate the structures, a median number of 10.5 steps is required per structure and the median energy difference between the structures before and after validation is 2.6 meV/atom.
Additionally, to demonstrate how the structure search is improved with an increasing number of active learning cycles, we deactivate the stopping criterion and continue the simulation for 10 active learning cycles.
The energies of the top 10 low-energy minima for all cycles are reported in the Supplementary Materials~\ref{app:si_all_cycles}.

\subsection{\label{subsec:results_transferability}Transferability to larger structures}

The computational cost of DFT calculations increases substantially with growing system size.
For complex systems with many atoms per periodic cell, even single-point evaluations for labeling training data can become computationally costly or infeasible.
Because the MLFFs use local atomic environments to model the structures, they are by design also transferable to larger periodic cells.
However, some local geometries of larger systems might not be captured by the training data and result in inaccurate predictions.
To test the transferability of the MLFFs towards systems with more atoms per cell, we perform simulations on crystal structures of the two different chemical compositions \ce{Si} and \ce{Al2O3}.

\begin{figure*}[!t]
  \centering
  \includegraphics[width=\textwidth]{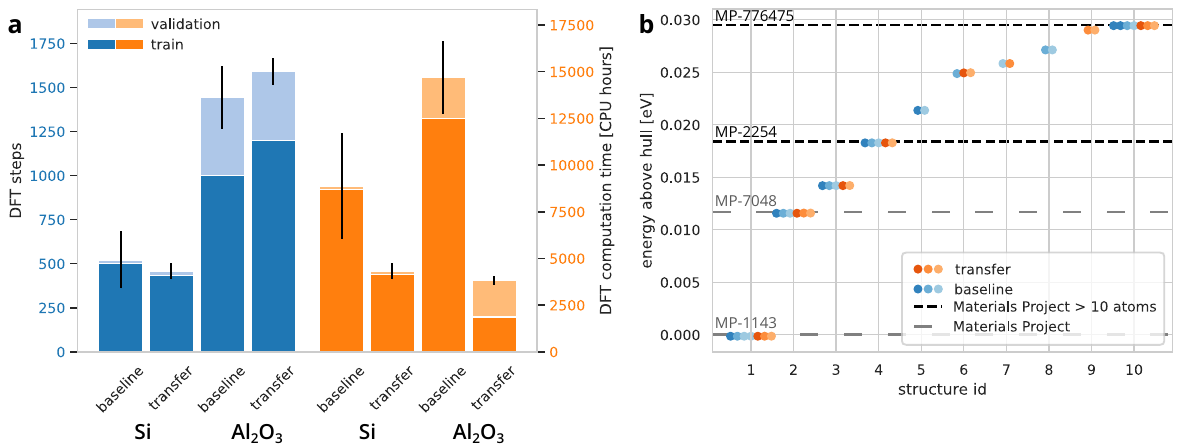}
  \caption{
  \textbf{a} Computational cost for global optimization simulations of \ce{Si} and \ce{Al2O3} using transfer learning compared to a baseline without transfer learning.
  For \ce{Si}, the transferability from the \ce{Si16} structures to the \ce{Si46} structures is tested, and for \ce{Al2O3}, the transferability is tested from \ce{Al4O6} to \ce{Al16O24}.
  The average number of DFT steps performed is shown by the blue bars, and the respective computation times are shown by the orange bars.
  The light shaded bars represent the computational cost for validating the proposed gloabal energy minimum, and the dark shaded bars measure computational cost during the active learning cycles.
  Every simulation is performed three times, and the errorbars show the standard deviation from the mean.
  \textbf{b} Comparison of energies for the top 10 low-energy clusters of \ce{Al16O24}.
  All energies of the structures proposed by the MLFFs are reported after validation through DFT structure relaxation.
  Orange dots are structures suggested by a model that has only trained on \ce{Al4O6} structures (transfer) and blue dots are suggested by MLFFs that have been trained on \ce{Al16O24} (baseline).
  Dashed lines show the energy levels of the \ce{Al2O3} structures with lowest energy provided in MaterialsProject, where black lines show the structures with more than 10 atoms per cell.
  For all models, the structures are sorted by their energy above the hull.
  }
  \label{fig:transferability}
\end{figure*}

For the simulations on \ce{Si}, we generate an initial pool of 100 \ce{Si46} structures per space group with symmetric generation that will be used to search for low-energy minima.
To sample training data for the MLFFs, we further generate a pool of 5000 randomly generated structures. 
In our transferability simulations, the training pool contains only \ce{Si16} structures, while our baseline experiment uses \ce{Si46} structures to sample training data, and during every active learning cycle.
As a target of our simulation, we use the lowest energy structure MP-971662 that is reported in MaterialsProject for \ce{Si46} structures.
The simulations on \ce{Al2O3}, which has a more complex PES than \ce{Si}, are initialized with a candidate pool of 500 \ce{Al16O24} structures per space group that we generate with symmetric generation.
To train the MLFFs, we generate 10,000 structures of \ce{Al4O6} for transferability simulations and 10,000 \ce{Al16O24} structures for baseline simulations, respectively.
We target MP-2254 as the \ce{Al16O24} structure with the lowest energy reported in MaterialsProject and we monitor the $k=10$ lowest energy clusters to determine convergence.
To account for stochasticity, we again perform three independent simulations for the baseline and transferability simulations of each chemical composition.
We compare the number of DFT steps that have been performed during the active learning cycles as well as during the validation of the global energy minimum to estimate the reduction in computational effort due to training on smaller structures.
Based on the average number of steps, the average computation times per step, and the number of 16 CPU cores that were used during the DFT computations, we estimate the total computation time of all DFT computations in CPU hours.
The number of steps and the computation times for all simulations of \ce{Si} and \ce{Al2O3} are shown in Figure~\ref{fig:transferability}~(left).

The simulations on \ce{Si} show that in both the transferability and the baseline simulation we are able to reliably find the target structure MP-971662.
For both cases this is achieved by selecting and validating the \ce{Si46} structure with the lowest energy prediction in the candidate pool after the stopping criterion has been triggered.
It is worth noting that MP-971661, the lowest energy structure of \ce{Si23}, is also found within the three lowest energy clusters of all simulations.
However, for measuring the computation times for the \ce{Si} simulations, we only count the validation of the global energy minimum.
The transferability simulations require approximately 4-5 cycles until convergence with about 500 single point DFT computations to find the global energy minimum, which is very similar to the baseline, but the computational cost is significantly smaller in the transferability simulations.
We estimate the computation time per DFT step by averaging all DFT steps that have been done during the iterative cycles and during validation, respectively, and we clip the maximum computation time per DFT step at a timeout of 100 minutes because failed DFT calculations otherwise distort estimate of the mean.
We further analyze the impact of timeout thresholds on the mean computation times in the Supplementary Materials~\ref{app:timeout}.
Using 16 CPUs, we measure an average computation time of 36 minutes for labeling a \ce{Si16} structure and 65 minutes for labeling a \ce{Si46} structure during active learning cycles.
During validation, the mean computation times per step are significantly lower at 28 minutes per step on \ce{Si46}, because the structures are already close to their local minimum and therefore do not suffer from timeouts or non-converging computations.
Multiplying the number of DFT steps by their average computation times yields a total computation time for all DFT steps of approximately 4322 CPU hours for the transferability simulation and approximately 8863 CPU hours for the baseline simulation, reducing the computational cost by 58~\%.

Transferability simulations in \ce{Al2O3} require approximately 1200 DFT calculations until the stopping criterion is triggered, which is slightly above the approximately 1000 steps that are performed in the baseline method. 
We measure an average computation time during the active learning cycles of 6 minutes per DFT calculation on \ce{Al4O6}, 47 minutes per DFT calculation on \ce{Al16O24} and an average computation time of 19 minutes during validation, so that the overall computational cost of all DFT calculations is reduced by approximately 74~\% due to transferability.

However, one out of three transferability simulations is not able to find the lowest energy structure MP-2254, while all baseline simulations do.
Figure~\ref{fig:transferability}~(right) shows the energy levels of the top 10 low-energy minima and the corresponding simulations that have found the local minimum, as well as the target structures of MaterialsProject for \ce{Al2O3}.
For the sake of clarity, we show only local minima that have been discovered by at least two different simulations.
The figure illustrates how all simulations find the structures MP-1143, MP-7048 and MP-776475, the second lowest energy minimum of \ce{Al16O24}.
The MP-2254 structure is found in all baseline simulations, but only two out of three transferability simulations are able to successfully discover it.
Furthermore, we also discover several local minima that are not included in MaterialsProject, but these minima are usually not discovered by all simulations.
Illustrations of all low-energy clusters from Figure~\ref{fig:transferability}~(right) can be found in the Supplementary Materials~\ref{app:transfer_Al2O3}.

\FloatBarrier
\section{\label{sec:conclusion}Discussion}

In this study, we introduced an iterative method for efficient discovery of low-energy crystal structures based on active learning with ensembles of neural network MLFFs.
We avoid up to 95~\% of the demanding DFT calculations in the relaxation trajectories by predicting the required physical properties with the trained MLFFs, compared to the baseline methods.
Besides speeding up structure relaxations, the MLFFs are also used to select promising structures for active learning, to extract representations for clustering the pool of candidates, and to provide a stopping criterion for our algorithm.
Thus, they enable automated, fast, and accurate global optimization of crystal structures.
For efficient computations on HPCs, computationally demanding tasks can be easily parallelized.
Since our uncertainty criterion is based on an MLFF ensemble, our method can be implemented with a plethora of state-of-the-art neural network architectures out of the box, providing the possibility for using pre-trained models or eventually foundation models that might arise in the near future.
We show that training neural network models on symmetric crystal structures significantly reduces the prediction accuracy, ultimately leading to MLFFs that cannot be used for reliable structure relaxations.
As remedies, we propose to create an additional training pool of randomly generated structures or to add small perturbations to selected structures before evaluation with DFT, which allows us to achieve the accuracy required for successful relaxation.

To evaluate the performance of our approach, we compared it with several approaches with full or partial structure relaxation trajectories on four different chemical compositions.
Our simulations show that we reduce the computational cost for finding the global energy minimum by up to two orders of magnitude.
Furthermore, because we simultaneously perform structure relaxation on the full candidate pool during every iteration, we demonstrate at the example of \ce{Si16} that our method can explore multiple low-energy structures while only adding minor computational effort.
Finally, we show that transferability from smaller cells that are computationally less expensive to evaluate towards large, challenging cells is possible, and can further reduce the computational cost.

Because our method allows a wide range of model architectures to be used as MLFFs, the computational cost could further be reduced in future work by starting the training of the MLFFs from pre-trained models.
Furthermore, extending our method with ML-based approaches for the initial candidate generation, as, for example, diffusion models, is another avenue to reduce the amount of DFT calculations for crystal structure search.
Finally, the computational cost of screening a single composition could be significantly reduced by simultaneously searching for low-energy structures of multiple chemical compositions while sharing the MLFFs.
Overall, we believe that our method will be broadly applicable as it constitutes a flexible framework for the accelerated exploration of crystal structures based on machine learning.
 
\section{\label{sec:methods}Methods}

\subsection{\label{subsec:training}Neural network models}
For all experiments, we use an ensemble of 5 SO(3)-equivariant neural networks based on spherical harmonics as implemented in SchnetPack \cite{schutt2019schnetpack, schutt2023schnetpack}.
After labeling new datapoints, the models are randomly initialized and trained independently on the updated database, excluding structures with forces that are larger than 3 standard-deviations above the mean.
Due to the comparably small number of training data, 2 interaction layers with a feature dimension of 64 are sufficient to accurately model the PES.
Furthermore, we allow for spherical harmonics features up to degree $l_{\text{max}}=2$ and use a cutoff of 7 $\mbox{\AA}$.
The models are trained using the Adam optimizer \cite{adam} with a batch size of 10 and we use an initial learning rate of $0.0005$. We monitor training progress on a 20 \% validation split and reduce the learning rate accordingly.
To handle the different magnitudes of the predicted values, we scale the MSE-loss (see Equation \eqref{eq:mse_loss}) with the weighting factors $\omega_{E}=0.0001$, $\omega_{f}=0.9999$, and $\omega_{s}=300$ for energy, forces, and stress, respectively.

\begin{equation}
\begin{split}
    \mathcal{L} = &\omega_{E} \cdot \norm{\edft-\epred}^2 \\
    &+ \omega_{f} \cdot \frac{1}{n_{\text{atoms}}}\sum_{i=1}^{n_{\text{atoms}}}\norm{\mathbf{f}_{\text{DFT, i}}-\mathbf{f}_{\theta, i}}^2 \\
    &+ \omega_{s} \cdot \frac{1}{9}\sum_{i, j=1}^{3}\norm{\mathbf{s}_{\text{DFT, ij}}-\mathbf{s}_{\theta, ij}}^2
\end{split}
    \label{eq:mse_loss}
\end{equation}

\subsection{\label{subsec:dft_settings}Reference computations}
All DFT computations are performed with the projector-augmented wave (PAW) method \cite{blochl1994projector, kresse1999ultrasoft}, using the implementation of Quantum Espresso \cite{QE-2009, QE-2017}.
We use pseudopotentials from PSlibrary \cite{dal2014pseudopotentials}, and the generalized gradient approximation (GGA) by Perdew, Burke, and Ernzerhof (PBE) \cite{perdew1996generalized} is employed for the exchange-correlation functional.
We use plane-wave cutoffs of 58 Ry, 66 Ry, 259 Ry and 323 Ry for \ce{Si}, \ce{NaCl}, \ce{GaAs} and \ce{Al2O3}, respectively.
The $k$-point mesh is dynamically generated with a mesh interval of 0.2 $\text{\AA}^{-1}$.
Non-converging calculations are stopped with a timeout of 2 h during all baseline simulations and 5 h during the transferability simulations.

\subsection{\label{subsec:structure_optimization}Structure relaxation}
For all structure relaxations, we use the ASE \cite{larsen2017atomic} implementation of the L-BFGS \cite{liu1989limited} optimizer with a force convergence threshold of 0.05 eV/$\mbox{\AA}$.
Furthermore, we simultaneously update the cell shape and the atomic positions during each step.
Structure relaxations based on property predictions of MLFFs use, in addition to the convergence threshold, an uncertainty stopping threshold according to equation \eqref{eq:uncertainty_crit} of $\uncert=0.5$.
To avoid large relaxation trajectories due to inaccuracies of the MLFFs, we stop structure relaxation if the energy has not decreased for more than 20 subsequent steps or if a maximum number of 300 steps is exceeded.

\subsection{\label{subsec:distance_constraints}Constraints for initial structure generation}
To ensure that the initial candidate structures remain within an acceptable range and to minimize the number of DFT calculations that do not converge, we establish constraints during the generation of structures.
If a structure violates any constraint, it is not added to the pool of candidates and a new structure is generated.
We specify $d_{ab}$ as the shortest permissable distance between two atoms of types $a$ and $b$, and the lengths of the lattice vectors are restricted to lie between $l_{\text{min}}$ and $l_{\text{max}}$.
We report the constraints that have been used during our simulations in Table \ref{tab:constraints}.
\begin{table}
    \centering
    \begin{tabular}{|c|c|c|c|c|}
         & $d_{aa}$ & $d_{ab}$ & $l_{\text{min}}$ & $l_{\text{max}}$ \\\hline
       \ce{Si16} & 1.6 & - & 3.0 & 15.0 \\
       \ce{Si46} & 1.6 & - & 4.0 & 22.0 \\
       \ce{Na8Cl8} & 2.0 & 1.5 & 3.0 & 15.0\\
       \ce{Ga8As8} & 2.0 & 1.5 & 3.0 & 15.0\\
       \ce{Al4O6} & 2.0 & 1.5 & 2.0 & 15.0\\
       \ce{Al4O6} (transfer) & 1.5 & 1.0 & 1.8 & 8.0\\
       \ce{Al16O24} (transfer) & 1.5 & 1.0 & 3.0 & 12.0\\
    \end{tabular}
    \caption{Constraints of minimal interatomic distances $d_{aa}$ and $d_{ab}$, and lattice boundaries $l_{\text{min}}$ and $l_{\text{max}}$ during structure generation. All values are reported in \AA.}
    \label{tab:constraints}
\end{table}
We observe that during the baseline simulations in Section \ref{subsec:results_transferability} none of the DFT calculations for the initial cycle of \ce{Al16O24} structures converges with the default constraints of \ce{Al4O6} from Section \ref{subsec:global_minimum}, which is why we update the constraints for \ce{Al2O3} for the transferability simulations.

\section*{Author contributions}
S.S.P.H. conceived and implemented the method, designed and conducted the experiments and drafted the manuscript.
K.T.S., N.W.A.G., and M.G. contributed to the development of the machine learning algorithms and the design of the experiments through insightful discussions and detailed analysis. 
T.O. and T.Y. provided essential assistance with analyzing and interpreting the crystal structures, designing the experiments, and setting up the DFT computations.
All authors contributed to the manuscript's development by offering critical revisions and insights during the research phase and reviewed the manuscript before submission.

\section*{Acknowledgements}
N.W.A.G. and M.G. contributed to this research while working at the BASLEARN—TU Berlin/BASF Joint Lab for Machine Learning, co-financed by TU Berlin and BASF SE.
K.T.S. contributed to this research while working at TU Berlin and BIFOLD with grant number 01IS18037A.
This work was also partly supported by JSPS KAKENHI Grant Number JP23H05457 and by JST-CREST Grant Number JPMJCR22O2.
We thank Jonas Lederer and Klaus-Robert Müller for insightful discussions and feedback.

\section*{Competing interests}
All authors declare no financial or non-financial competing interests.

\section*{Data availability}
The datasets generated and analyzed during this study are publicly available in the DepositeOnce repository with the DOI: 10.14279/depositonce-21008~\cite{hessmann2024simdata}.
This includes the initial and final candidate pools, validated structures, generated training data, and trained neural network ensembles from the last iteration.
Researchers can access and utilize these datasets for further studies or replication efforts.
Additional data is provided upon reasonable request.

\section*{Code availability}
The code supporting the findings of this study is openly available on GitHub at \url{https://github.com/stefaanhessmann/active-csp} (DOI 10.5281/zenodo.13145078~\cite{hessmann2024activecsp}). This repository includes instructions for installation and usage, along with the configuration files that are used to generate the results.

\bibliography{references}

\clearpage

\section{\label{app}Supplementary Materials}

\subsection{Training performance on symmetric structures\label{app:mlff_symmetry}}
\begin{figure}
    \centering
    \includegraphics[width=0.5\textwidth]{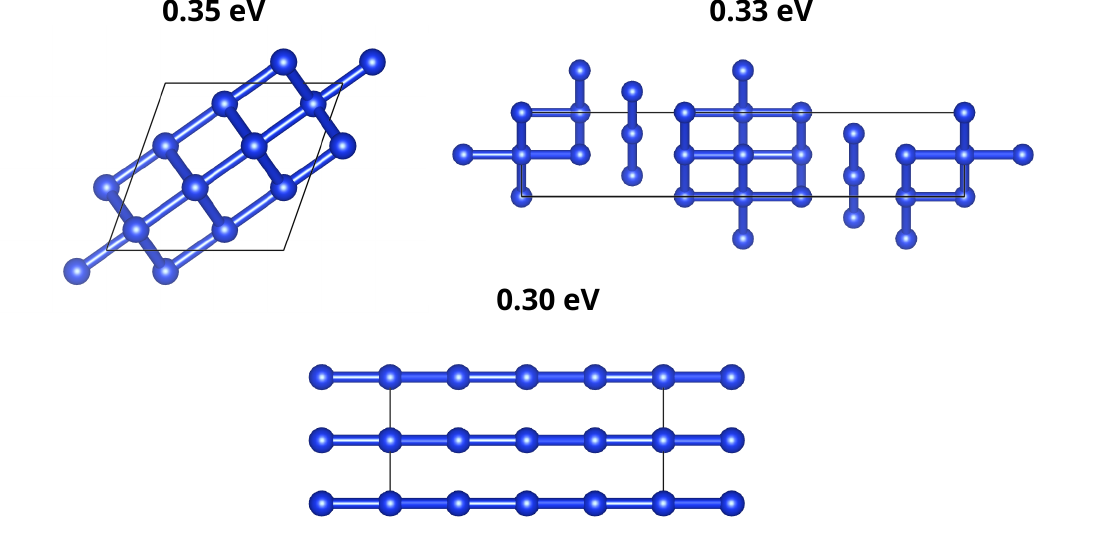}
    \caption{
    Top \ce{Si16} structures from symmetric structure simulation. The renders of the top 3 \ce{Si16} structures are found by a simulation that is trained on symmetric structures without adding noise.
    Energy values show the average energy per atom above the complex hull of \ce{Si16}.
    }
    \label{fig:nonoise_structures}
\end{figure}

As Cheon \textit{et al.}~\cite{cheon2020dataset} show, training neural network MLFFs on structure relaxation trajectories that start with symmetric crystal structures significantly decreases the prediction performance of the trained models.
In our simulations, we also experience a decrease in the prediction performance of models that have been trained solely on symmetric crystal structures.
To test this, we run three independent simulations simulation on \ce{Si16} with the same settings as in Section \ref{subsec:global_minimum}, but only use an initial pool of candidates that was generated through symmetric generation and do not add noise $\noiset$ to the selected structures before labeling. 
We find that none of the simulations is able to find the global energy minimum after the stopping criterion has been triggered.
Instead, often structures with very simple symmetries with repeating local atomic neighborhoods, such as rectangular and triangular patterns, are modeled accurately, but more complex structures are not sufficiently optimized during structure relaxation.
Figure \ref{fig:nonoise_structures} shows representatives, validated with DFT relaxation, of the top three low-energy clusters that our method selects after the stopping criterion is triggered.
Clearly, the MLFFs that are trained on symmetric structures preferably select structures with simple geometries, although these local minima are far above the complex hull of \ce{Si16}. 

\subsection{\label{app:runtime}Runtime evaluation}
\begin{figure}
    \centering
    \includegraphics[width=0.5\textwidth]{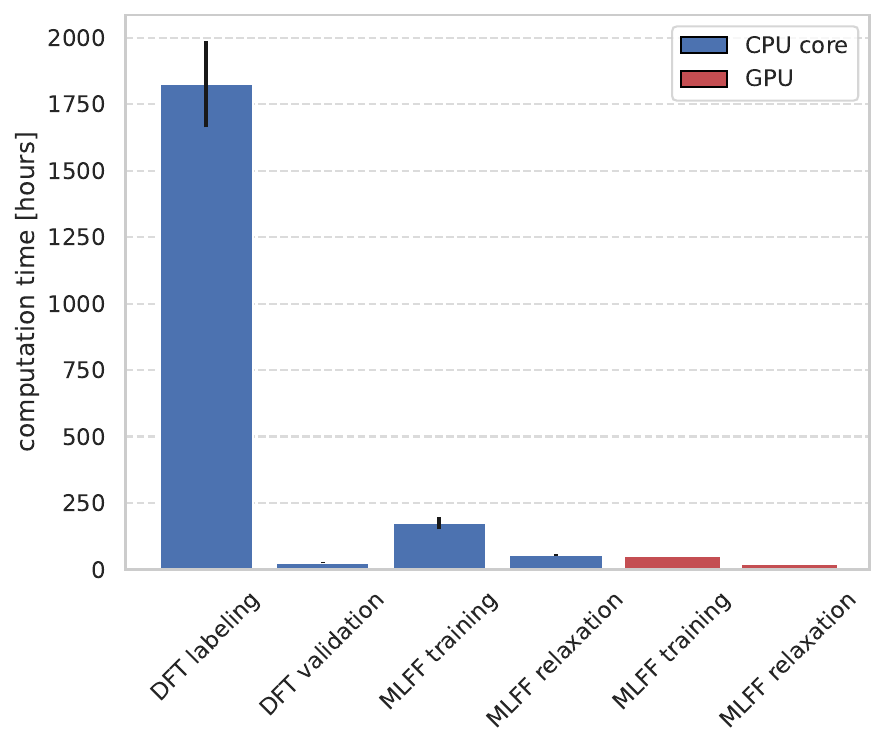}
    \caption{Average computation times for the different stages of our method during simulations on \ce{Si16}. The measured computation times are multiplied by the number of devices that have been used. All DFT computations are performed on 16 CPU cores and the MLFF runs use 4 CPU cores and 1 GPU. Error bars show the standard deviation from the mean of 3 simulations.}
    \label{fig:ctimes}
\end{figure}

Compared to the other approaches in Section \ref{subsec:global_minimum} the MLFFs introduce additional computational cost during the active learning cycles.
To account for this, we evaluate the computation times for a complete simulation at the example of the global minimum search of \ce{Si16}.
Besides the DFT computations for labeling training data and validating the low-energy structures, we identify the training of MLFFs and using them for structure relaxation of the large candidate pools as the main computationally expensive tasks.
To estimate the overall computational cost of our method, we measure the runtime of every task and accumulate the computational cost per device over the full simulation.
All DFT computations are performed on a device with 16 CPU cores, while training and property prediction with neural networks are done with a single GPU and 4 CPU cores.
The average computation times per CPU and GPU device of all four stages are shown in Figure \ref{fig:ctimes}.

The measurements show that about 90\% of all CPU core hours are performed during labeling the training data with DFT.
Training the MLFFs requires the second most CPU computation time, but still remains negligible in comparison.
For our experiments, we select 100 structures per cycle for labeling with DFT and use an ensemble size of 5 neural networks.
Furthermore, all neural networks are trained on separate devices and structure relaxation with the MLFFs is done sequentially.
By training the neural network models on the same device and by performing batch wise structure relaxation of the candidate pool, the computational cost could significantly be further reduced.
However, because the computational cost that is introduced by active learning during these simulations is comparably small, this is neglected in this work, but could help to speed up the screening of significantly larger candidate pools.

\subsection{Top 10 \ce{Si16} low-energy structures\label{app:si_minima}}
\begin{figure*}[!t]
  \centering
  \includegraphics[width=\textwidth]{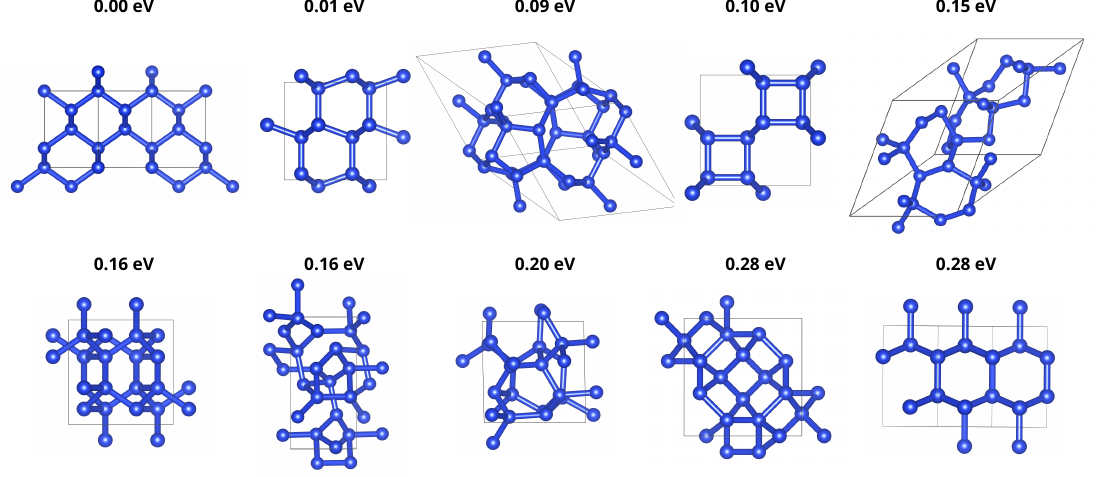}
  \caption{
  Example low-energy \ce{Si16} structures of a multiple minima search simulation.
  All structures have been validated by structure relaxation with DFT.
  Structures are sorted according to DFT energies, and the energy values show the energy per atom about the complex hull.
  }
  \label{fig:si16_vesta}
\end{figure*}

Figure \ref{fig:si16_vesta} shows the stable structures that are observed in the simulation for finding multiple low-energy minima for \ce{Si16} (see Section \ref{subsec:multi_search}).
We use the VESTA software to create graphic illustrations of the 3d structures and select projections that capture the most significant features of the crystal structure.
To obtain these structures, a simulation with a stopping criterion of $k=10$ is performed until convergence and the most promising low-energy structures that our method proposed are validated through structures relaxation with DFT.

\subsection{Evaluation of low-energy \ce{Si16} structures for all cycles\label{app:si_all_cycles}}
\begin{figure*}[!t]
  \centering
  \includegraphics[width=\textwidth]{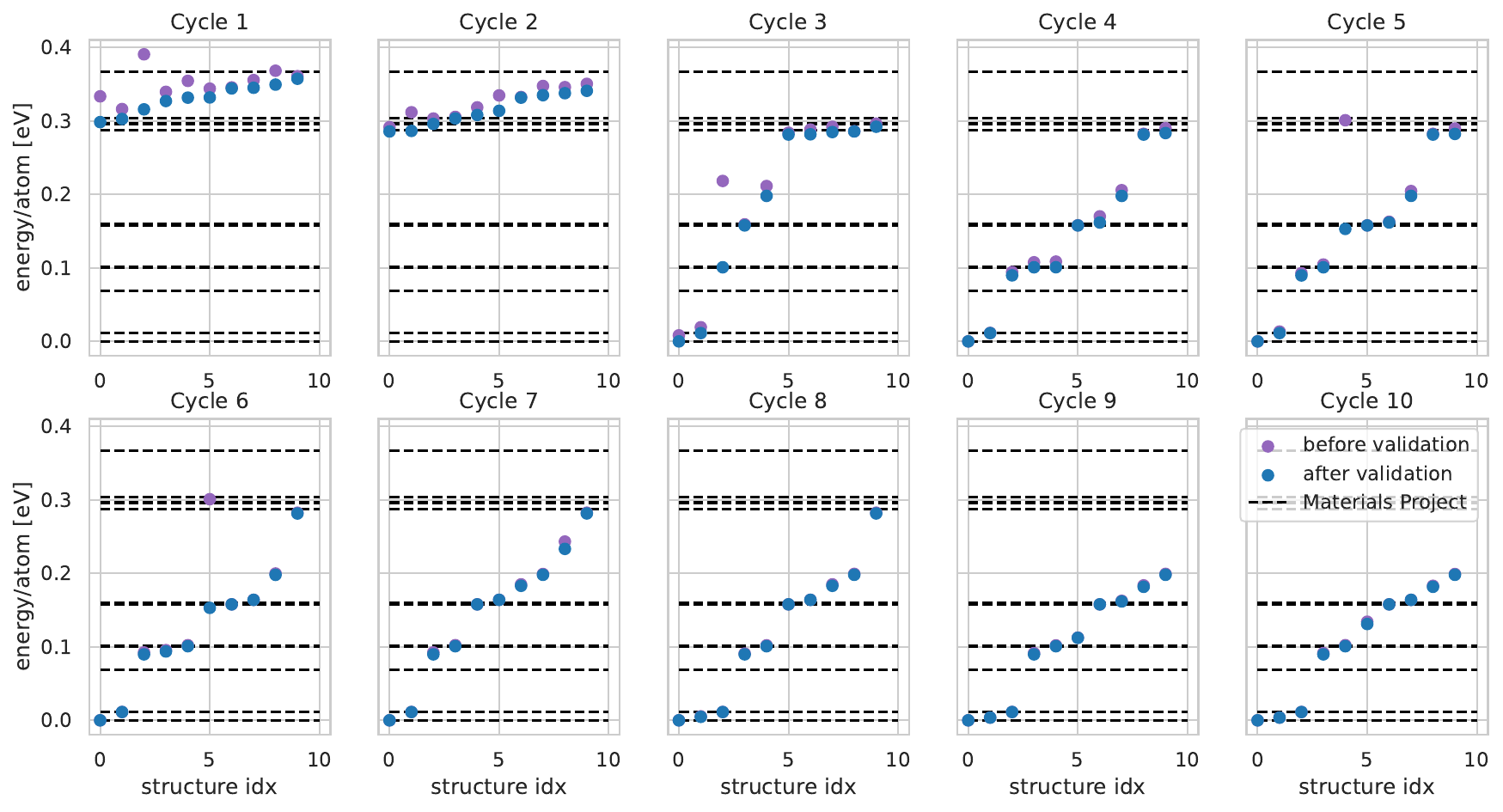}
  \caption{
  Energies of top 10 proposed low-energy clusters of \ce{Si16} during 10 active learning cycles.
  Structures before validation are shown with purple dots, and structures after validation by DFT are shown in blue.
  Dashed black lines show energies of the respective low-energy structures as reported in MaterialsProject.
  Low-energy structures, found by our method, are sorted according to their energy after validation through DFT.
  }
  \label{fig:10_cycles}
\end{figure*}

For a deeper understanding how the MLFFs and the candidate pool are improved during the active learning cycles, we extend the simulation of finding multiple \ce{Si16} low-energy minima (see Section \ref{subsec:multi_search}) to run for 10 full cycles, without using the stopping criterion.
At the end of every cycle, we validate the 10 lowest energy structures that are selected by the MLFFs through structure relaxation with DFT.
In Figure \ref{fig:10_cycles}, we show the DFT energies of selected low-energy structures before and after validation, as well as the energies of the lowest 10 structures reported by the MaterialsProject database for \ce{Si16} structures.

During the first two cycles the DFT energies of the proposed low-energy structures are still far above the complex hull, but with an increasing number of cycles the energies of selected structures are significantly reduced.
After only 300 DFT calculations that are used to label training data for the first three cycles, the global energy minimum is included in the proposed low-energy structures and on average the proposed structures have lower energies than the reference structures of MaterialsProject.
This trend continues until the energy levels start to converge.
Please note that because we validate the selected structures with DFT structure relaxation, all local minima of this simulation are stable structures with respect to the reference DFT calculations.
In addition to the energies of the proposed minima, also the energy difference between structures before and after validation is decreased with every epoch, indicating more accurate estimates of the low-energy minima.
While in the first cycle, the energy difference between the structures before and after validation is approximately 0.02 eV/atom, the difference is reduced to 0.001 eV/atom at the last cycle.

\subsection{DFT computation times\label{app:timeout}}

\begin{figure}
    \centering
    \includegraphics[width=0.5\textwidth]{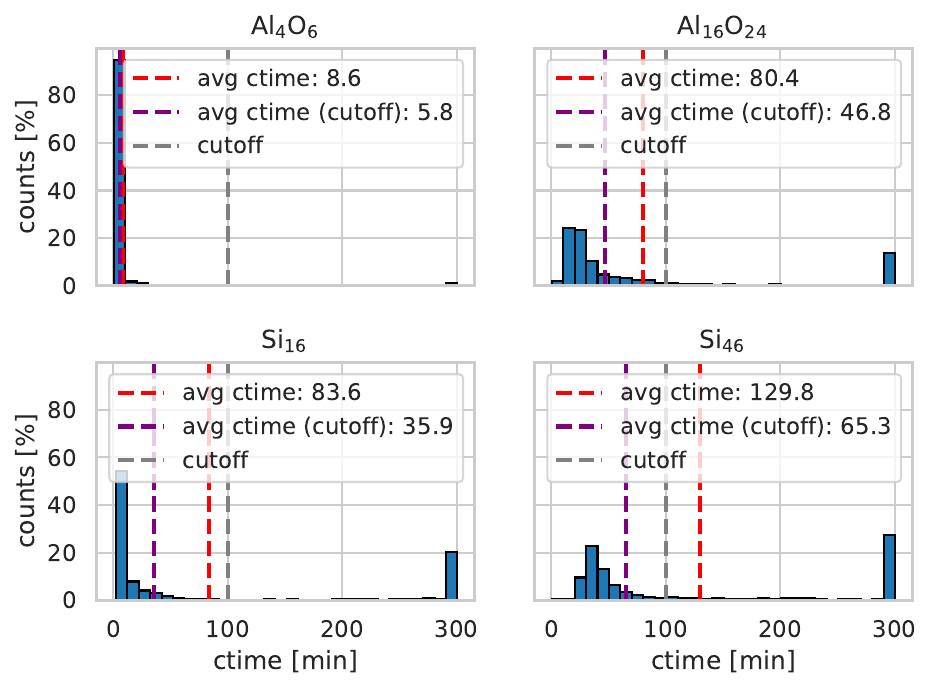}
    \caption{
    Histogram of computation times per DFT step.
    All DFT steps are performed during the active learning cycles of the transferability experiments for \ce{Si} and \ce{Al2O3}. All computations are performed on a device with 16 CPU cores with a timeout threshold of 300 minutes. Dashed lines show the mean computation times per DFT calculation with the timeout threshold of 300 minutes (red), a timeout threshold of 100 minutes (purple) and the reduced timeout threshold of 100 minutes (grey).
    }
    \label{fig:transfer_timings_hist}
\end{figure}

DFT calculations can, depending on the input structure, fail to converge.
To keep the overall computation times in a reasonable range, we set a threshold of 300 minutes per DFT step that defines the calculation as a failure case.
However, when analyzing the average computational cost of DFT calculations, we find that the failed calculations significantly distort the estimates without any contribution to training the MLFFs.
Figure \ref{fig:transfer_timings_hist} shows the histograms of all DFT computations that have been performed during the active learning cycles of the transferability experiments in Section \ref{subsec:results_transferability}.
The histograms show that apart from the failed computations at the 300 min threshold, nearly no other computations required more than 100 min until convergence.
Still, the imbalanced distribution results in significantly overestimated computation times due to the failure cases.
To account for these failure cases, we virtually set the timeout value in the comparison of computation times between baseline and transferability simulations to 100 min.

\subsection{Top 10 \ce{Al16O24} low-energy structures\label{app:transfer_Al2O3}}

\begin{figure*}[!t]
  \centering
  \includegraphics[width=\textwidth]{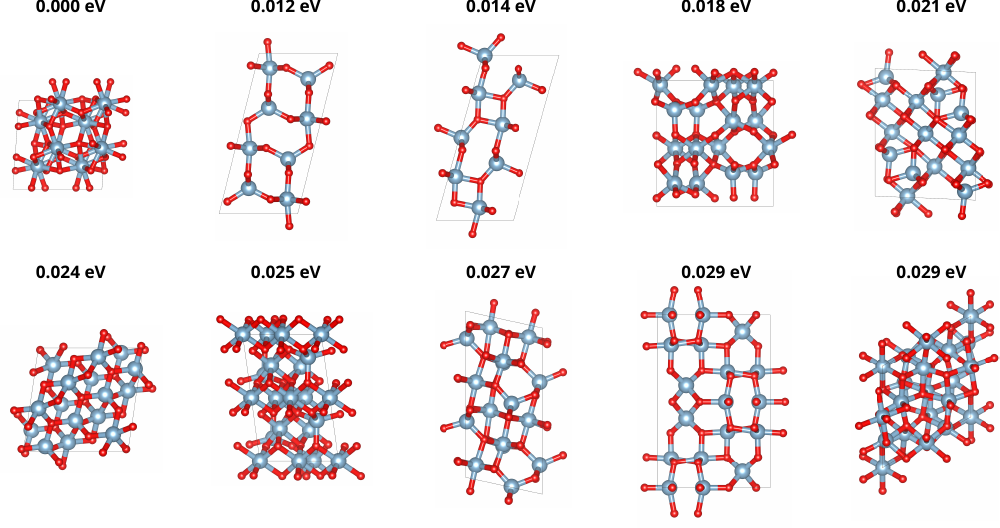}
  \caption{
  Example low-energy \ce{Al16O24} structures.
  Structures are sorted according to DFT energies, and the energy values show the energy per atom about the complex hull.
  }
  \label{fig:alo_vesta}
\end{figure*}

In the simulations of Section \ref{subsec:results_transferability} we show that the baseline method as well as the transferability simulations find several local minima in the PES of \ce{Al2O3}.
While some of those structures are reported in MaterialsProject, our method also finds additional stable structures that we validated through structure relaxation with DFT.
Figure \ref{fig:alo_vesta} shows the illustrations of those structures generated with VESTA and their corresponding atomic energy above the complex hull.
The lowest two energy minima that we find with all of the simulations require up to 10 atoms per cell, and they correspond to the structures MP-1143 and MP-7048 of MaterialsProject.
Next, the structure at 0.014 eV/atom is also found with all simulations, and it shows a very similar pattern to the energetically similar structure MP-7048, but according to the DFT validation, it is a separated local minimum on its own.
The structures at 0.018 eV/atom and 0.029 eV/atom match the structures MP-2254 and MP-776475 of MaterialsProject, which require 40 atoms per cell, and only one of the transferability simulations fails to correctly identify the structure MP-2254. 
Additionally, in the range of 0.021 eV/atom to 0.029 eV/atom, some of our simulations find additional low-energy minima that are not reported in MaterialsProject but fulfill the stability requirements of the DFT relaxation that is performed during validation.
Furthermore, we also identify several stable low-energy minima that are only found by a single simulation, but for clarity we have omitted those structures in the analysis.
However, for a deeper understanding of those rare local minima that are not reported in MaterialsProject, additional simulations with stricter settings for the parameters $n_{\text{cycle}}$ and $k$ are needed.

\end{document}